\documentclass[aps,prb,reprint,superscriptaddress,showpacs,showkeys,amsmath,floatfix,longbibliography]{revtex4-1}

\usepackage{color}
\usepackage{graphicx}
\usepackage{dcolumn}
\usepackage{bm}
\usepackage{textcomp}
\usepackage{verbatim}
\usepackage{upgreek}
\usepackage{comment}
\usepackage{hyperref}

\newcommand{\nv}{68}
\newcommand{\Lambdaafit}{107}
\newcommand{\Lambdaa}{100}

\newcommand{\Lambdaafite}{33}
\newcommand{\Lambdaae}{35}

\newcommand{\Lambdacfit}{134}
\newcommand{\Lambdac}{120}
\newcommand{\Lambdaclow}{102}
\newcommand{\Lambdacfite}{6}
\newcommand{\Lambdace}{15}
\newcommand{\Lambdaclowe}{15}

\newcommand{\gammaL}{1.2}

\newcommand{\gammaLe}{0.4}
\newcommand{\xiVfit}{31}
\newcommand{\xiV}{36}
\newcommand{\xiVfite}{9}
\newcommand{\xiVe}{10}

\newcommand{\xia}{41}

\newcommand{\xiae}{13}

\newcommand{\xic}{34}

\newcommand{\xice}{10}

\begin{document}


\title{Measurement of the Penetration Depth and Coherence Length of
  MgB$_2$ in All Directions Using Transmission Electron Microscopy}

\author{J.C. Loudon}
\email{j.c.loudon@gmail.com}
\affiliation{Department of Materials Science and
  Metallurgy, 27 Charles Babbage Road, Cambridge, CB3 0FS, United Kingdom}

\author{S. Yazdi}

\author{T. Kasama}
\affiliation{Centre for Electron Nanoscopy, Technical University of Denmark, DK-2800 Kongens Lyngby, Denmark}

\author{N.D. Zhigadlo}
\affiliation{Laboratory for Solid State Physics, ETH Zurich, Otto-Stern-Weg 1, CH-8093, Zurich, Switzerland}

\author{J. Karpinski}
\affiliation{Laboratory for Solid State Physics, ETH Zurich, Otto-Stern-Weg 1, CH-8093, Zurich, Switzerland}
\affiliation{Institute of Condensed Matter Physics, EPFL, 1015-Lausanne, Switzerland}

\date{\today}


\begin{abstract}

We demonstrate that images of flux vortices in a superconductor taken
with a transmission electron microscope can be used to measure the
penetration depth and coherence length in all directions at the same
temperature and magnetic field. This is particularly useful for
MgB$_2$, where these quantities vary with the applied magnetic field
and values are difficult to obtain at low field or in the $c$
direction. We obtained images of flux vortices from a MgB$_2$ single
crystal cut in the $ac$ plane by focussed ion beam milling and tilted
to $45^\circ$ with respect to the electron beam about the
crystallographic $a$ axis. A new method was developed to simulate
these images which accounted for vortices with a non-zero core in a
thin, anisotropic superconductor and a simplex algorithm was used to
make a quantitative comparison between the images and simulations to
measure the penetration depths and coherence lengths. This gave
penetration depths $\Lambda_{ab}=\Lambdaa\pm\Lambdaae$~nm and
$\Lambda_c=\Lambdac\pm\Lambdace$~nm at 10.8~K in a field of
4.8~mT. The large error in $\Lambda_{ab}$ is a consequence of tilting
the sample about $a$ and had it been tilted about $c$, the errors on
$\Lambda_{ab}$ and $\Lambda_c$ would be reversed. Thus, obtaining the
most precise values requires taking images of the flux lattice with
the sample tilted in more than one direction. In a previous paper
[J. C. Loudon {\it et al.}, Phys. Rev. B 87, 144515, 2013], we obtained a more precise value for
$\Lambda_{ab}$ using a sample cut in the $ab$ plane. Using this value
gives $\Lambda_{ab}=107\pm 8$~nm, $\Lambda_c=\Lambdac\pm\Lambdace$~nm,
$\xi_{ab}=39\pm 11$~nm and $\xi_c=35\pm 10$~nm which agree well with
measurements made using other techniques. The experiment required two
days to conduct and does not require large-scale facilities. It was
performed on a very small sample: $30\times 15$~$\upmu$m and 200~nm
thick so this method could prove useful for superconductors where only
small single crystals are available, as is the case for some
iron-based superconductors.

\end{abstract}



\maketitle


\section{Introduction}
\label{Introduction}

Superconductors have zero electrical resistance and expel magnetic
flux from their interiors (the Meissner effect). However, if a
sufficiently high magnetic field is applied, flux penetrates by
flowing along channels called flux vortices. Each vortex carries one
quantum of magnetic flux, $\Phi_0=h/2e$ where $h$ is Planck's constant
and $e$ the electron charge. They consist of a core with a radius
given by the coherence length, $\xi$, where the number of carriers
(electrons or holes) contributing to superconductivity is
suppressed. Electrical supercurrents circulate around the centre,
diminishing over a radius given by the penetration depth,
$\Lambda$. In a conventional superconductor, the coherence length is
related to the energy required to excite a carrier out of the
superconducting state, $\Delta$, and the velocity of the carriers at
the Fermi energy, $v_F$, via $\xi=\hbar v_F/\pi\Delta$ and the
penetration depth is related to the number density of carriers
involved in superconductivity, $n_S$, and their effective mass, $m^*$,
via $\Lambda=\sqrt{m^*/\mu_0 n_S e^2}$ ($\mu_0$ is the permeability of
free space).

In a type-I superconductor, the core exceeds the size over which the
supercurrents persist and vortices attract one another as the area of
normal (non-superconducting) material is minimised if the cores
overlap. In a type-II superconductor, the supercurrents persist over a
larger radius than the core and the Lorentz force causes vortices to
repel one another so they form a hexagonal array in an isotropic
superconductor. Introducing the Ginzburg-Landau parameter,
$\kappa\equiv \Lambda/\xi$: a type-I superconductor has
$\kappa<1/\sqrt{2}$ and type-II has $\kappa>1/\sqrt{2}$.

\begin{figure}
\includegraphics[width=80mm]{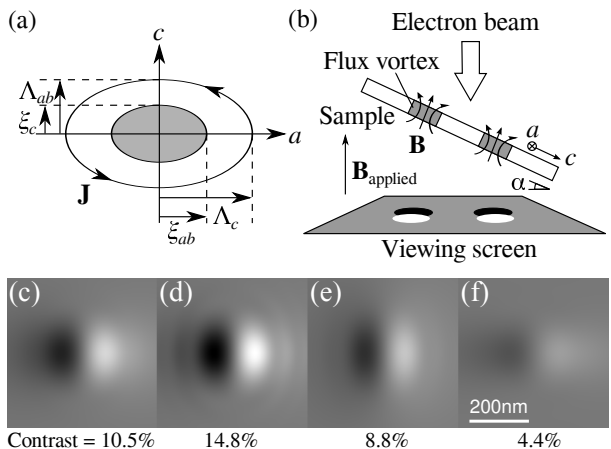}
\caption{\label{geometry} (a) Flux vortex with its axis normal to the
  $ac$ plane. The core (shaded) is elliptical with dimensions
  $\xi_{ab}$ and $\xi_c$. Supercurrents ${\bf J}$ follow ellipses with
  the same aspect ratio, diminishing on lengthscales $\Lambda_c$ and
  $\Lambda_{ab}$. Streamlines of ${\bf J}$ are also contours of
  magnetic flux density ${\bf B}$. (b) Experimental arrangement for
  imaging flux vortices. The electron beam is deflected by the
  component of ${\bf B}$ normal to the beam so vortices appear as
  black-white features in an out-of-focus image. (c)--(f) Simulated
  images with defocus $\Delta f=7.15$~mm for a flux vortex in a 180~nm
  thick specimen in the orientation shown in (a) but tilted $45^\circ$
  about $a$. Contrast values (see text) are shown below each
  image. (c) $\Lambda_{ab}=100$~nm, $\Lambda_c=120$~nm, $\xi_V=34$~nm,
  (d) $\Lambda_{ab}=100$~nm, $\Lambda_c=120$~nm, $\xi_V=1$~nm, (e)
  $\Lambda_{ab}=200$~nm, $\Lambda_c=120$~nm, $\xi_V=34$~nm, (f)
  $\Lambda_{ab}=100$~nm, $\Lambda_c=200$~nm, $\xi_V=34$~nm.}
\end{figure}

An anisotropic superconductor has different properties along different
crystal axes, $a$, $b$ and $c$. Most are uniaxial so that $a$ and $b$
are equivalent. The anisotropy in the penetration depth is
$\gamma_\Lambda\equiv\Lambda_c/\Lambda_{ab}$ and in the coherence
length it is $\gamma_\xi\equiv\xi_{ab}/\xi_c$.  In a 1-band
superconductor, where there is one source of carriers contributing to
superconductivity, the penetration depth and coherence length are
independent of the applied magnetic field and their anisotropies are
equal. One method to investigate the penetration depths and coherence
lengths in both the $a$ and $c$ directions is to induce flux vortices
with their axes normal to the $ac$ plane. The vortex then has an
elliptical core surrounded by circulating currents as illustrated in
Fig.~\ref{geometry}(a).

MgB$_2$ is a rare 2-band superconductor~\cite{Nagamatsu01} discovered
in 2001 with a transition temperature $T_c=39$~K. It is uniaxial with
a hexagonal crystal structure~\cite{Nagamatsu01} (space group 191:
$P6/mmm$) composed of alternating layers of magnesium and boron with
lattice parameters $a=b=3.086$~\AA~and $c=3.542$~\AA. Two bands
contribute to superconductivity: the $\sigma$-band associated with
bonding from the boron $p_{xy}$ orbitals and the $\pi$-band associated
with boron $p_z$ orbitals~\cite{Choi02}. $\sigma$ carriers are
confined to the $ab$ planes whereas the $\pi$ carriers are delocalised
almost isotropically. At low magnetic fields, both $\sigma$ and $\pi$
bands contribute to superconductivity but as the field is increased,
the $\pi$ contribution diminishes so that above 0.8~T (at 2~K), only
the $\sigma$ band contributes~\cite{Cubitt03}. This has the effect
that the penetration depth and coherence length vary with
field~\cite{Cubitt03}.

In a previous paper~\cite{Loudon13}, we showed that the penetration
depth of MgB$_2$, $\Lambda_{ab}$, could be obtained in the low-field
limit by making a quantitative comparison between images of flux
vortices acquired using transmission electron microscopy and
simulations. Here, we extend this method and show that the penetration
depths $\Lambda_{ab}$ and $\Lambda_c$ and coherence lengths $\xi_{ab}$
and $\xi_c$ can be measured in a low field of 4.8~mT from a very small
sample.

Focussed ion beam milling was used to cut a MgB$_2$ sample in the $ac$
plane of size $30\times 15$~$\upmu$m, thinned to 200~nm so that it was
electron transparent (see section~\ref{methods}). Flux vortices
penetrate normal to the thin surfaces and the sample was tilted about
its $a$ axis at $\alpha=45\pm 5^\circ$ to give a component of the
B-field normal to the electron beam (Fig.~\ref{geometry}(b)). The beam
is deflected by the Lorentz force and flux vortices appear as
black-white features in an out-of-focus image~\cite{Harada92}. Such
images are sensitive to the $B$-field throughout the thickness of the
specimen, not just the stray field, and they can be acquired in real
time~\cite{youtube}.

The effect on such images of changing the coherence length and
penetration depth is shown in Fig.~\ref{geometry}(c)--(f). These
images were simulated by extending Beleggia's method~\cite{Beleggia02}
to model vortices with a non-zero core in a thin, anisotropic
superconductor (see section~\ref{simulation}). We use Klemm and Clem's
Ginzburg-Landau model for the vortex core~\cite{Clem80, Klemm80}. In
this model, the core has the same symmetry as the circulating currents
so that $\xi_{ab}/\xi_c=\Lambda_c/\Lambda_{ab}$ and
$\xi_V=\xi_{ab}^{1/3}\xi_c^{2/3}$ but any model for the magnetic
structure of a flux vortex could be used with equal facility.

Fig.~\ref{geometry}(c) shows a simulated image of a vortex with
$\Lambda_{ab}=100$~nm, $\Lambda_c=120$~nm and $\xi_V=34$~nm. (d) shows
that decreasing $\xi_V$ to 1~nm sharpens the image, increasing the
contrast (the difference in the maximum and minimum intensities
divided by their sum) from 10.5\% to 14.8\%. In (e), $\Lambda_{ab}$ is
doubled which stretches the image in $c$ and reduces the contrast from
10.5\% to 8.8\%. (f) shows the images are most sensitive to
$\Lambda_c$ so that when $\Lambda_c$ is increased to 200~nm, the image
is stretched in $a$ and its contrast falls to 4.4\%. This sensitivity
of the images to changes in these parameters should allow the
simultaneous measurement of the penetration depths and coherence
lengths in all directions. In this paper we assess the accuracy of
this new technique.


\section{Simulation of Flux Vortex Images}
\label{simulation}

In this section, we present a model to calculate the magnetic fields
generated by a flux vortex and from this simulate transmission
electron micrographs. The model accounts not only for the B-field
inside the superconductor but also for the spreading of the field
lines near the superconductor surface and the field outside. It
extends Beleggia's method~\cite{Beleggia01,Beleggia02} to treat the
case of a vortex in a thin, anisotropic superconducting slab and makes
use of the work of Klemm and Clem~\cite{Clem80, Klemm80, Clem92} to
account for a non-zero vortex core although it has the convenient
feature that any model for the vortex core can be used with equal
facility. Like all magnetic objects, flux vortices change only the
phase and not the intensity of the electron beam and once the fields
have been calculated, the phase shift can be found using the
Aharanov-Bohm formula. Once the phase shift is known, any image can be
simulated. Thus we first evaluate the magnetic vector potential, then
use this to find the phase shift and from this simulate out-of-focus
images of flux vortices.

\subsection{Coordinate Systems}

\begin{figure}
\includegraphics[width=50mm]{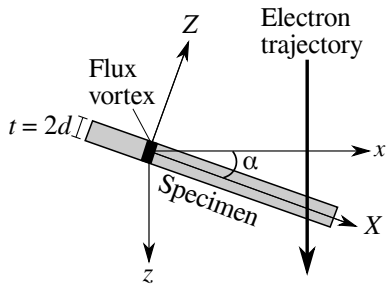}
\caption{\label{coordinates} The relationship between the coordinates
  $X,Y,Z$, referring to the specimen and the microscope coordinates
  $x,y,z$. The $y$ and $Y$ axes are normal to the other two axes and
  point in the direction given by the right-hand rule. The axis of the
  vortex is parallel to $Z$ and the B-field enters the specimen at the
  bottom and exits at the top. The electron beam is parallel to $z$.}
\end{figure}

In order to visualise flux vortices using transmission electron
microscopy, the specimen must be tilted by an angle $\alpha$ to
provide a component of the B-field normal to the electron beam so that
the electrons are deflected by the Lorentz force and show contrast in
an out-of-focus image. Thus we follow Beleggia's
method~\cite{Beleggia01,Beleggia02} and introduce two coordinate
systems: $X,Y,Z$ referring to the specimen with $X$ and $Y$ in the
specimen plane and $Z$ normal to its surface and $x,y,z$ referring to
the microscope with $z$ parallel to the electron beam as illustrated
in Fig.~\ref{coordinates}. The specimen surfaces are at $Z=\pm d$ so
its thickness is $t=2d$. We first evaluate the magnetic vector
potential in terms of the specimen coordinates and use this to find
the phase shift in the $xy$ plane which is equivalent to the plane in
which images are recorded.

\subsection{Magnetic Vector Potential}

Here we evaluate the magnetic vector potential, ${\bf A}$, from a flux
vortex passing through a thin superconducting slab with its axis
directed along $Z$, normal to its surfaces. We use Beleggia's
Fourier-space method~\cite{Beleggia01} throughout in which all the
functions are Fourier transformed in $x$ and $y$ but not $z$ (or $X$
and $Y$ but not $Z$). This allows the Fourier transform of the vector
potential and phase to be expressed by analytical but very lengthy
expressions. These were evaluated symbolically using Matlab and only
the final inverse transforms were performed numerically. For both
coordinate systems, we use the transform convention that if $g(x,y,z)$
is a function in real-space, its Fourier transform
$\widetilde{g}(k_x,k_y,z)$ is:

\begin{equation}
\widetilde{g}(k_x,k_y,z)=\int\limits_{-\infty}^\infty \int\limits_{-\infty}^\infty{g(x,y,z)e^{-i(k_xx+k_yy)}\,{\rm d}x{\rm d}y}
\end{equation}

If the order parameter of the superconducting state is written in
terms of its amplitude $f$ and phase $\theta$ as $\Psi=fe^{i\theta}$,
the vector potential inside the superconductor is related to it by the
2nd Ginzburg-Landau equation~\cite{Annett04, Ginzburg50}:

\begin{equation}
\label{GL2}
{\bf A}f^2+L\nabla\times(\nabla\times {\bf A})=\frac{\Phi_0}{2\pi}f^2\nabla\theta
\end{equation}

where $L$ is penetration depth tensor. When $X,Y,Z$ are principal axes
of the superconductor it has components:

\begin{equation}
L=\left(\begin{array}{ccc}
\Lambda_{_X}^2 & 0 & 0 \\
0 & \Lambda_{_Y}^2 & 0 \\
0 & 0 & \Lambda_{_Z}^2
\end{array}\right)
\end{equation}

Following Clem~\cite{Clem80}, we look for a solution of the form

\begin{equation}
{\bf A}={\bf A}_\text{bulk}+{\bf A}_\text{surface}
\end{equation}

${\bf A}_\text{bulk}$ is the solution for a single vortex in an
infinitely thick specimen and consequently has no
$Z$-dependence. ${\bf A}_\text{surface}$ is the general solution with
the correct boundary conditions but with the right-hand side of the
Ginzburg-Landau equation set to zero.

Clem~\cite{Clem75} solved for the bulk term in the isotropic case by
using an order parameter for a single vortex of the form

\begin{equation}
\Psi=fe^{i\theta}=\frac{\rho}{\sqrt{\rho^2+\xi^2}}\,e^{-i\phi}
\end{equation}

where $\rho$ is the radius from the axis of the vortex, $\phi$ is the
azimuthal angle and $\xi$ is the coherence length. Klemm and
Clem~\cite{Klemm80,Clem92} later extended this solution to the anisotropic
case so that the order parameter becomes:

\begin{equation}
\Psi\!=\!\sqrt{\frac{X^2/\Lambda_{_Y}^2+Y^2/\Lambda_{_X}^2}{X^2/\Lambda_{_Y}^2+Y^2/\Lambda_{_X}^2+\xi^2/\Lambda^2}}\,e^{-i\arg(X/\Lambda_{_Y}+iY/\Lambda_{_X})}
\end{equation}

where $\Lambda=(\Lambda_{_X}\Lambda_{_Y}\Lambda_{_Z})^{1/3}$. This gives the
magnetic flux density as

\begin{equation}
{\bf B}=\frac{\Phi_0}{2\pi\Lambda_{_X}\Lambda_{_Y}}\frac{K_0(R)}{(\xi/\Lambda)K_1(\xi/\Lambda)}\widehat{\bf Z}
\end{equation}

where $R=\sqrt{(X/\Lambda_{_Y})^2+(Y/\Lambda_{_X})^2+(\xi/\Lambda)^2}$ and
$K_0$ and $K_1$ are zero and first order modified Bessel functions.

Fourier transforming the flux density gives

\begin{equation}
\widetilde{\bf B}=\frac{\Phi_0K_1(Q\xi/\Lambda)}{Q K_1(\xi/\Lambda)}\widehat{\bf Z}
\end{equation}

where $Q=\sqrt{1+k_{_X}^2\Lambda_{_Y}^2+k_{_Y}^2\Lambda_{_X}^2}$. Since ${\bf B}=\nabla\times{\bf A}$, in Fourier space we have $\widetilde{\bf B}=(ik_{_X}\widetilde{A}_Y-ik_{_Y}\widetilde{A}_X)\widehat{\bf Z}=i{\bf k}_\perp\times\widetilde{\bf A}$ where ${\bf k}_\perp=(k_{_X},k_{_Y},0)$. Imposing the additional requirement that ${\bf A}$ obey the London gauge $\nabla.{\bf A}=0$ so that $ik_{_X}\widetilde{A}_X+ik_{_Y}\widetilde{A}_Y=0$, the vector potential in Fourier space is

\begin{equation}
\widetilde{\bf A}_\text{bulk}=
\frac{i\Phi_0K_1(Q\xi/\Lambda)}{k_\perp^2Q K_1(\xi/\Lambda)}\left(\begin{array}{c}
k_{_Y}\\
-k_{_X}\\
0
\end{array}\right)
\end{equation}

We now find the surface term, ${\bf A}_{\rm surface}$. This is the
general solution to the Ginzburg-Landau equation but where the
right-hand side is set equal to zero. We use the approximation
introduced by Clem~\cite{Clem80} that the surface term should be only
weakly influenced by the vortex core and so set $f^2=1$ in the search
for a solution. The validity of this simplification was confirmed by
Brandt~\cite{Brandt04} who modelled the complete vortex and found that
the core only expands by a few percent as it approaches the surface of
the superconductor. The surface term is thus the solution to

\begin{equation}
\label{surface}
{\bf A}+L\nabla\times(\nabla\times {\bf A})=0
\end{equation}

It should be noted that setting the right hand side of the
Ginzburg-Landau equation (eqn.~\ref{GL2}) to zero fixes the gauge of
the vector potential and for an anisotropic superconductor, this is
not the London gauge. Thus we cannot say $\nabla{\bf .A}=0$ nor make
the convenient replacement $\nabla\times(\nabla\times {\bf
  A})=-\nabla^2{\bf A}$. Instead we must deal with the awkward
cross-terms arising from the double curl.

Taking the Fourier transform of Eqn.~\ref{surface} gives:

\begin{equation}
\widetilde{\bf A}+L(i\,{\bf k}_\perp+\widehat{\bf Z}\,\partial_{_Z})\times((i\,{\bf k}_\perp+\widehat{\bf Z}\,\partial_{_Z})\times \widetilde{\bf A})=0
\end{equation}

We now postulate a solution of the form $\widetilde{\bf
  A}(k_{_X},k_{_Y},Z)={\bf a}(k_{_X},k_{_Y})e^{\beta Z}$ and the resulting
equation can be written in matrix form as:

\begin{widetext}
\begin{equation}
\left(\begin{array}{ccc}
1+k_{_Y}^2\Lambda_{_X}^2-\Lambda_{_X}^2\beta^2 & -k_{_X}k_{_Y}\Lambda_{_X}^2 & ik_{_X}\Lambda_{_X}^2\beta \\
-k_{_X}k_{_Y}\Lambda_{_Y}^2 & 1+k_{_X}^2\Lambda_{_Y}^2-\Lambda_{_Y}^2\beta^2 & ik_{_Y}\Lambda_{_Y}^2\beta \\
ik_{_X}\Lambda_{_Z}^2\beta & ik_{_Y}\Lambda_{_Z}^2\beta & 1+\Lambda_{_Z}^2(k_{_X}^2+k_{_Y}^2)

\end{array}\right)
{\bf a}=\left(\begin{array}{c} 0 \\ 0 \\ 0\end{array}\right)
\end{equation}
\end{widetext}

The above equation gives non-zero solutions for the vector potential
if the matrix cannot be inverted. To achieve this, values of $\beta$
must be found to make the determinant zero. At this point we introduce
the symmetry of the problem otherwise the answers become very
lengthy. For this experiment, the specimen was tilted about the $a$
axis and thus, $X \parallel c$, $Y \parallel a$ and $Z \parallel b$ so
that $\Lambda_{_X}=\Lambda_c$ and $\Lambda_{_Y}=\Lambda_{ab}$ and
$\Lambda_{_Z}=\Lambda_{ab}$. There are then four possible values of
$\beta$:

\begin{equation}
\beta_{1,3}=\pm Q_a/\Lambda_{ab}
\end{equation}

and

\begin{equation}
\beta_{2,4}=\pm Q/\Lambda_{c}
\end{equation}

with corresponding eigenvectors

\begin{equation}
{\bf a}_{1,3}=\left(\begin{array}{c}
0 \\
\pm i Q_a/(k_{_Y} \Lambda_{ab}) \\
1
\end{array}\right)
\end{equation}

and

\begin{equation}
{\bf a}_{2,4}=\left(\begin{array}{c}
\pm i \Lambda_c(1+k_{_X}^2\Lambda_{ab}^2)/(\Lambda_{ab}^2 k_{_X} Q) \\
\pm i \Lambda_ck_{_Y}/Q  \\
1
\end{array}\right)
\end{equation}

where $Q_a=\sqrt{1+(k_{_X}^2+k_{_Y}^2)\Lambda_{ab}^2}$ and
$Q=\sqrt{1+k_{_X}^2\Lambda_{ab}+k_{_Y}^2\Lambda_c}$. The complete vector
potential inside the superconductor is then:

\begin{equation}
\widetilde{\bf A}_\text{inside}=\widetilde{\bf A}_\text{bulk}+\sum_{n=1}^4 c_n{\bf a}_ne^{\beta_n z}
\end{equation}

where $c_1$ -- $c_4$ need to be determined by the boundary conditions. 

This leaves the vector potential outside the superconductor to be
determined. Maxwell's third equation gives $\nabla\times {\bf B}=0$ as
there are no electrical currents outside the superconductor so the
vector potential obeys $\nabla\times(\nabla\times {\bf A})=0$. This
time, the London gauge, $\nabla{\bf .A}=0$, may safely be invoked to
give

\begin{equation}
\nabla^2{\bf A}_\text{outside}=0
\end{equation}

or, in Fourier space:

\begin{equation}
-k_\perp^2\widetilde{\bf A}+\frac{\partial^2 \widetilde{\bf A}}{\partial Z^2}=0
\end{equation}

The solution to this is

\begin{equation}
\widetilde{\bf A}_\text{top}=\left(\begin{array}{c}
c_5 \\
c_6 \\
i(c_5k_{_X}+c_6k_{_Y})/k_\perp
\end{array}\right)
e^{-k_\perp Z}
\end{equation}

\begin{equation}
\widetilde{\bf A}_\text{bottom}=\left(\begin{array}{c}
c_7 \\
c_8 \\
-i(c_7k_{_X}+c_8k_{_Y})/k_\perp
\end{array}\right)
e^{k_\perp Z}
\end{equation}

where the $Z$-components are determined by the London gauge.

We can now simplify the equations as symmetry requires that
$A_{_{X,Y}}(-Z)=A_{_{X,Y}}(Z)$ and $A_{_Z}(-Z)=-A_{_Z}(Z)$. This gives
$c_1=-c_3$, $c_2=-c_4$, $c_5=c_7$ and $c_6=c_8$.

Summarising so far, the vector potential inside the superconductor is

\begin{eqnarray}
\label{inside}
\widetilde{\bf A}_\text{inside}=\widetilde{\bf A}_\text{bulk}
&+&2c_1\left(\begin{array}{c}
a_{_{1,X}}\cosh(\beta_1 Z) \\
a_{_{1,Y}}\cosh(\beta_1 Z) \\
a_{_{1,Z}}\sinh(\beta_1 Z)
\end{array}\right)\nonumber \\
&+&2c_2\left(\begin{array}{c}
a_{_{2,X}}\cosh(\beta_2 Z) \\
a_{_{2,Y}}\cosh(\beta_2 Z) \\
a_{_{2,Z}}\sinh(\beta_2 Z)
\end{array}\right)
\end{eqnarray}

and above and below the superconductor, the vector potential is

\begin{equation}
\label{outside}
\widetilde{\bf A}_\text{above, below}=\left(\begin{array}{c}
c_5 \\
c_6 \\
\pm i(c_5k_{_X}+c_6k_{_Y})/k_\perp
\end{array}\right)
e^{\mp k_\perp Z}
\end{equation}

To fix the values of $c_1$, $c_2$, $c_5$ and $c_6$, we invoke the
following boundary conditions: (1) In order to calculate the phase
shift from the vector potential, the $X$ and $Y$ components of the
vector potential must change continuously across the boundary between
the superconductor and vacuum at $Z=\pm d$. (2) The in-plane flux
density $B_\parallel$ must be continuous across the boundaries at
$Z=\pm d$ as there are no currents confined to the surface of the
superconductor. There is also the requirement that the normal
component of the flux density $B_\perp$ be continuous but this arises
from Maxwell's 3rd equation, $\nabla.{\bf B}=0$, and by using a vector
potential, it is automatically satisfied.

Condition (1) that the in-plane vector potential is continuous at
$Z=\pm d$ gives two equations (one for each component):

\begin{eqnarray}
\widetilde{A}_{_{\text{bulk},X}}
+2c_1 a_{_{1,X}}\cosh(\beta_1 d)\nonumber
+2c_2 a_{_{2,X}}\cosh(\beta_2 d)\\
=c_5e^{-k_\perp d}
\end{eqnarray}

\begin{eqnarray}
\widetilde{A}_{_{\text{bulk},Y}}
+2c_1 a_{_{1,Y}}\cosh(\beta_1 d)\nonumber
+2c_2 a_{_{2,Y}}\cosh(\beta_2 d)\\
=c_6e^{-k_\perp d}
\end{eqnarray}

Calculating the flux density via ${\bf B}=\nabla\times {\bf A}$ or, in
Fourier space, $\widetilde{\bf B}=(i{\bf k}_\perp+\widehat{\bf
  z}\partial_z)\times \widetilde{\bf A}$ and matching its in-plane
components at the interface gives two more:

\begin{eqnarray}
(-(k_{_Y}/k_\perp)(c_5k_{_X}+c_6k_{_Y})+c_6k_\perp)e^{-k_\perp d}=\nonumber \\
2c_1D_1\sinh(\beta_1 d)+2c_2D_2\sinh(\beta_2 d)
\end{eqnarray}

\begin{eqnarray}
(-c_5k_\perp+(k_{_X}/k_\perp)(c_5k_{_X}+c_6k_{_Y}))e^{-k_\perp d}=\nonumber \\
2c_1D_3\sinh(\beta_1 d)+2c_2D_4\sinh(\beta_2 d)
\end{eqnarray}

where $D_1=ik_{_Y}a_{_{1,Z}}-\beta_1a_{_{1,Y}}$,
$D_2=ik_{_Y}a_{_{2,Z}}-\beta_2a_{_{2,Y}}$, $D_3=\beta_1a_{_{1,X}}-ik_{_X}a_{_{1,Z}}$
and $D_4=\beta_2a_{_{2,X}}-ik_{_X}a_{_{2,Z}}$

Writing these four equations in matrix form gives:

\begin{widetext}
\begin{equation}
\label{final}
\left(\begin{array}{cccc}
2D_1\sinh(\beta_1d) & 2D_2\sinh(\beta_2d) & \frac{k_{_X}k_{_Y}}{k_\perp}e^{-k_\perp d} & \left(\frac{k_{_Y}^2}{k_\perp}-k_\perp\right)e^{-k_\perp d} \\
2D_3\sinh(\beta_1d) & 2D_4\sinh(\beta_2d) & -\left(\frac{k_{_X}^2}{k_\perp}-k_\perp\right)e^{-k_\perp d} & -\frac{k_{_X}k_{_Y}}{k_\perp}e^{-k_\perp d} \\
2a_{1,x}\cosh(\beta_1d) & 2a_{2,x}\cosh(\beta_2d) & -e^{-k_\perp d} & 0 \\
2a_{1,y}\cosh(\beta_1d) & 2a_{2,y}\cosh(\beta_2d) & 0 & -e^{-k_\perp d}
\end{array}\right)
\left(\begin{array}{c}
c_1 \\
c_2 \\
c_5 \\
c_6
\end{array}\right)=
\left(\begin{array}{c}
0 \\
0 \\
-\widetilde{A}_{_{\text{bulk},X}} \\
-\widetilde{A}_{_{\text{bulk},Y}}
\end{array}\right)
\end{equation}
\end{widetext}

The coefficients $c_1$, $c_2$, $c_5$, $c_6$ can then be found by
inverting the matrix and then, after substituting the answers into
Eqns.~\ref{inside} and \ref{outside}, the vector potential is fully
determined. It should be noted that although we use Klemm and Clem's
solution for ${\bf A}_{\text{bulk}}$, Eqn.~\ref{final} shows that our method
has the convenient feature that any model for ${\bf A}_{\text{bulk}}$ could
be used with equal facility.


\subsection{Phase Shift}

\begin{figure}
\includegraphics[width=40mm]{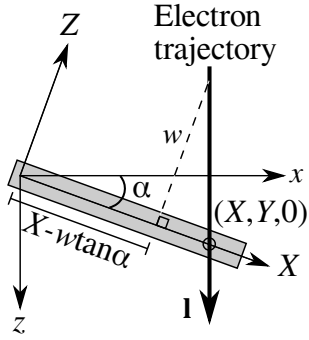}
\caption{\label{w} The relationship between the coordinates $X,Y,Z$,
  referring to the specimen (shown by the grey rectangle) and the
  microscope coordinates $x,y,z$.}
\end{figure}

The magnetic contribution to the phase shift suffered by the electron
beam after passing through a specimen is related to the vector
potential via the Aharanov-Bohm expression:

\begin{equation}
\phi(x,y)=-\frac{2\pi e}{h}\int_{-\infty}^\infty {\bf A}(x,y,z){\bf .}{\rm d}{\bf l}
\end{equation}

where ${\rm d}{\bf l}$ is an increment along the trajectory of the
electrons shown in Fig.~\ref{w}. If unit vectors in $x$, $y$ and $z$
of the microscope coordinate system are denoted ${\bf i}$, ${\bf j}$
and ${\bf k}$ and those in $X$, $Y$ and $Z$ of the specimen coordinate
system are denoted ${\bf I}$, ${\bf J}$ and ${\bf K}$, the two sets of
unit vectors are related via:

\begin{eqnarray}
{\bf I}&=&{\bf i}\cos\alpha+{\bf k}\sin\alpha \\
{\bf J}&=&-{\bf j} \\
{\bf K}&=&{\bf i}\sin\alpha-{\bf k}\cos\alpha
\end{eqnarray}

If the electron passes through the specimen at position $(X,Y,0)$
and if for the preceding part of its journey, we label the height
above the specimen in the $Z$ direction $w$ (see Fig.~\ref{w}),
its position at any point in its trajectory is given by

\begin{equation}
{\bf l}=(X-w\tan\alpha){\bf I}+Y{\bf J}+w{\bf K}
\end{equation}

By differentiating the above equation, an increment in its trajectory
${\rm d}{\bf l}$ can be related to an increment in $w$ via:

\begin{equation}
{\rm d}{\bf l}=(-\tan\alpha\,{\bf I}+{\bf K}){\rm d}w
\end{equation}

The phase shift written in terms of the specimen coordinates is thus:

\begin{eqnarray}
&&\phi(X,Y)=-\frac{2\pi e}{h}\times\nonumber\\
&&\int_{-\infty}^\infty {\bf A}(X-w\tan\alpha,Y,w).\left(\begin{array}{c} -\tan\alpha \\ 0 \\ 1 \end{array}\right)\,{\rm d}w
\end{eqnarray}

In the previous section the vector potential was calculated in Fourier
space. This is related to the vector potential in real-space via the
inverse Fourier transform:

\begin{eqnarray}
{\bf A}(&X&,Y,Z)=\nonumber \\
&&\frac{1}{4\pi^2}\int\limits_{-\infty}^\infty\int\limits_{-\infty}^\infty{\widetilde{\bf A}(k_{_X},k_{_Y},Z)e^{i{\bf k}_\perp.{\bf R}}\,{\rm d}k_{_X}\,{\rm d}k_{_Y}}
\end{eqnarray}

where ${\bf R}=(X,Y,Z)$. Thus it follows that

\begin{eqnarray}
&&{\bf A}(X-w\tan\alpha,Y,Z)=\nonumber \\
&&\frac{1}{4\pi^2}\iint\widetilde{\bf A}(k_{_X},k_{_Y},Z)e^{i{\bf k}_\perp.{\bf R}}e^{-ik_{_X}w\tan\alpha}\,{\rm d}k_{_X}\,{\rm d}k_{_Y}
\end{eqnarray}

\begin{widetext}

So the phase shift is

\begin{equation}
\phi(X,Y)=-\frac{2\pi e}{h}\int_{-\infty}^\infty 
\left(\frac{1}{4\pi^2}\iint \widetilde{\bf A}(k_{_X},k_{_Y},w)e^{i{\bf k}_\perp.{\bf R}}e^{-ik_{_X}w\tan\alpha}\,{\rm d}k_{_X}\,{\rm d}k_{_Y}\right)
.\left(\begin{array}{c} -\tan\alpha \\ 0 \\ 1 \end{array}\right)\,{\rm d}w
\end{equation}

It can be seen from Fig.~\ref{w} that if an electron passes through a
point $(X,Y,0)$ on the specimen, it passed through a point $(x,y,0)$
in the $xy$ plane where $x=X\cos\alpha$ and $y=-Y$. The phase shift in
the $xy$-plane (which is equivalent to the plane in which the image is
taken) is now found by making this substitution.

\begin{equation}
\phi(x,y)=-\frac{2\pi e}{h}\int_{-\infty}^\infty 
\left(\frac{1}{4\pi^2}\iint \widetilde{\bf A}(k_{_X},k_{_Y},w)e^{ik_{_X}x/\cos\alpha}e^{-ik_{_Y}y}e^{-ik_{_X}w\tan\alpha}\,{\rm d}k_{_X}\,{\rm d}k_{_Y}\right)
.\left(\begin{array}{c} -\tan\alpha \\ 0 \\ 1 \end{array}\right)\,{\rm d}w
\end{equation}

Now let $k_x=k_{_X}/\cos\alpha$ and $k_y=-k_{_Y}$:

\begin{equation}
\phi(x,y)=\frac{2\pi e}{h}\int_{-\infty}^\infty 
\left(\frac{1}{4\pi^2}\iint \widetilde{\bf A}(k_x\cos\alpha,-k_y,w)e^{ik_xx}e^{ik_yy}e^{-ik_xw\sin\alpha}\cos\alpha\,{\rm d}k_x\,{\rm d}k_y\right)
.\left(\begin{array}{c} -\tan\alpha \\ 0 \\ 1 \end{array}\right)\,{\rm d}w
\end{equation}

Changing the order of integration gives:

\begin{equation}
\phi(x,y)=\frac{2\pi e\cos\alpha}{h}\frac{1}{4\pi^2}
\iint e^{i{\bf k_\perp.r}}\left(\int_{-\infty}^\infty \widetilde{\bf A}(k_x\cos\alpha,-k_y,w)e^{-ik_xw\sin\alpha}\,{\rm d}w\right)\,{\rm d}k_x\,{\rm d}k_y
.\left(\begin{array}{c} -\tan\alpha \\ 0 \\ 1 \end{array}\right)
\end{equation}

or representing an inverse Fourier transform as IFT and using the flux
quantum $\Phi_0$ we have:

\begin{equation}
\phi(x,y)={\rm IFT}\left[\frac{\pi}{\Phi_0}\left(\begin{array}{c} -\sin\alpha \\ 0 \\ \cos\alpha \end{array}\right).\,
\int_{-\infty}^\infty \widetilde{\bf A}(k_x\cos\alpha,-k_y,w)e^{-ik_xw\sin\alpha}\,{\rm d}w\right]
\end{equation}

\end{widetext}

The integral over $w$ is straightforward as it only involves
exponential functions but it is very lengthy so it and the scalar
product were performed symbolically using Matlab. Only the final
inverse Fourier transform which gives the phase was evaluated numerically.

\begin{figure}
\includegraphics[width=84mm]{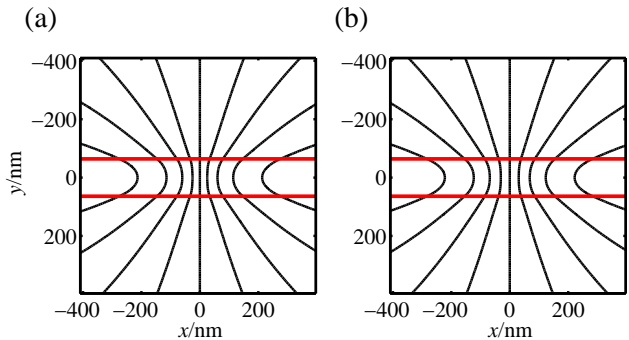}
\caption{\label{B_field} Contours of the phase-shift spaced by 0.2~rad
  illustrating the projected B-field from a flux vortex with
  $\Lambda_{ab}=100$~nm and $\Lambda_c=120$~nm and (a) $\xi_V=1$~nm and (b)
  $\xi_V=36$~nm. The red lines indicate the surfaces of the specimen which
  is 128~nm thick. The $a$-axis is parallel to $x$ with $c$ pointing
  into the page.}
\end{figure}

A method to check the correctness of the solution is to plot the phase
shift as a contour map for $\alpha=90^\circ$. This gives the B-field
projected through the thickness of the specimen and is shown in
Fig.~\ref{B_field}. It can be seen that the correct boundary
conditions are fulfilled: the field lines spread as they approach the
specimen surface from the interior and outside the specimen they are
straight so that the field resembles that from a monopole if viewed
far from the vortex. The figure shows that the effect of increasing
the coherence length $\xi_V$ is to make the field less intense near the
centre of the vortex as expected.

\subsection{Image Simulation}

Once the phase shift $\phi(x,y)$ has been calculated, the wavefunction
of the electron beam is $\psi_0(x,y)=e^{i\phi}$ and the intensity of the
in-focus bright-field image is $I_0(x,y)=|\psi_0(x,y)|^2$. It is
immediately clear that this gives 1 and an in-focus image is therefore
featureless. In order to visualise flux vortices, out-of-focus images
must be taken. Taking an out-of-focus image is equivalent to
propagating the wavefunction through free space by a distance $\Delta
f$, known as the defocus. This is done via the Fresnel-Kirchoff
integral~\cite{Beleggia02} so that the defocussed wavefunction
$\psi_{\Delta f}(x,y)$ is related to the in-focus wavefunction via:

\begin{eqnarray}
\psi_{\Delta f}(x,y)=&&\nonumber\\
\frac{1}{\lambda\Delta f}\iint&\psi_0(x',y')&e^{\frac{i\pi}{\lambda\Delta f}((x-x')^2+(y-y')^2)}\,{\rm d}x'{\rm d}y'
\end{eqnarray}

where $\lambda$ is the electron wavelength. This is a convolution and
so is more conveniently evaluated as a multiplication in Fourier space
via:

\begin{equation}
\widetilde\psi_{\Delta f}=\widetilde\psi_0 e^{-i\lambda\Delta f k^2/4\pi}
\end{equation}

After inverse transforming, the intensity in the out-of-focus image is
given by $I_{\Delta f}=|\psi_{\Delta f}|^2$.


\section{Experimental Method}
\label{methods}

MgB$_2$ single crystals were synthesised by the peritectic
decomposition of MgNB$_9$ and their quality and bulk properties have
been well characterised by a variety of experimental
techniques~\cite{Karpinski03,Karpinski07,Zhigadlo10}. The samples were
prepared for electron microscopy at the Technical University of
Denmark (DTU) using a Helios Nanolab focussed ion beam microscope
(FIB). This is a dual-beam instrument in which a beam of gallium ions
is used to mill the specimen whilst secondary electrons emitted by the
specimen are used to produce an image, however an electron beam can
also be used to illuminate the specimen in order to take images
without damaging the specimen.

The MgB$_2$ single crystals were about $1 \times 1$~mm in the $ab$
plane and about 100~$\upmu$m thick in the $c$-direction. The {\it in-situ}
lift-out technique was used to prepare the sample for electron
microscopy. First, the FIB was used to deposit a 3~$\upmu$m thick,
$30\times 5$~$\upmu$m rectangle of platinum onto the $ab$ surface to
protect the sample beneath from ion damage. Trenches were then milled
to a depth of $20$~$\upmu$m in the $c$ direction around this to
produce a slab standing in the centre of a crater. The top surface of
the sample was smooth and this avoided the creation of the
longitudinal thickness undulations reported in our last
paper~\cite{Loudon13}.

A movable needle known as a micromanipulator was attached to the slab
using platinum deposition and the slab was cut away from the rest of
the specimen and extracted on the end of the micromanipulator. A
sample can be tilted only to $25^\circ$ in the electron microscope so
to achieve a higher tilt angle, the FIB was used to cut a slot at
$45^\circ$ to the plane of an `Omniprobe' grid. Further platinum
deposition was used to attach the sample to this slot and the
micromanipulator was then cut away, leaving the sample attached to the
grid and tilted about its $a$ axis by $45^\circ$ with respect to the
plane of the grid. The sample was then thinned to approximately 200~nm
so that it was electron-transparent using a 30~kV Ga ion
beam. Finally, the specimen surfaces were polished by a low-energy
(2kV) Ga ion beam to minimise the damage layer caused by FIB milling.

Electron microscopy was undertaken at DTU using an FEI Titan 80-300ST
transmission electron microscope operated at 300~kV equipped with a
Gatan imaging filter to record images. Under normal operating
conditions, the main objective lens of the microscope applies a 2~T
field to the specimen so to avoid this, the microscope was operated in
low-magnification mode with the main objective lens set to a low value
and the image was focussed with the diffraction lens. Prior to imaging
vortices, electron diffraction was used to make a fine adjustment of a
few degrees so that the tilt was purely about the $a$ axis. Adjusting
the sample so that it is tilted purely about $a$ can be performed to
better than $0.5^\circ$ but this may alter the overall tilt angle and
we judge that the tilt angle was $\alpha=45\pm 5^\circ$.

The simulations were based on elastic electron scattering so
experimental images were energy filtered so that only electrons which
had lost 0--10~eV on passing through the specimen contributed and an
aperture was used so that only the 000 beam and the low-angle
scattering from the vortices contributed to the image and the other
crystallographic beams were excluded. The sample was cooled using a
Gatan liquid-helium cooled `IKHCHDT3010-Special' tilt-rotate holder
which has a base temperature of 10~K.

The defocus and magnification were calibrated by acquiring images with
the same lens settings as the original images from Agar Scientific's
`S106' calibration specimen which consists of lines spaced by 463~nm
ruled on an amorphous film. The defocus was found by taking digital
Fourier transforms of the images acquired from the calibration
specimen and measuring the radii of the dark rings which result from
the contrast transfer function~\cite{Williams96}.

A thickness map of the specimen was created by dividing an unfiltered
image by an energy-filtered image and taking the natural logarithm
\cite{Egerton09} which gives the thickness parallel to the electron
beam, $l$, as a multiple of the inelastic mean free path,
$\lambda_i$. To determine $\lambda_i$, an electron hologram was taken
at room temperature at an edge of the specimen which gives a phase
shift proportional to the thickness, $\phi=C_EV_0l$. $C_E$ is a
constant which depends only on the microscope voltage and has the
value $6.523 \times 10^6$~m$^{-1}$V$^{-1}$ at 300~kV. $V_0$, the mean
inner potential, was calculated as $V_0=17.71$~V from theoretical
scattering factors given in ref.~\onlinecite{Rez94}, giving
$\lambda_i=244\pm 5$~nm and the thickness, $l$, varied from
200--290~nm across the field of view of Fig.~\ref{defseries}. Ideally
the thickness of the whole specimen would have been determined by
electron holography but the field of view of the interference region
was not sufficiently large.

A simplex algorithm~\cite{Press92} was used to minimise the reduced
$\chi^2$ value between the experimental images and simulations by
fitting the vortex positions and the in-plane rotation angle of the
vortices as well as $\Lambda_{ab}$, $\Lambda_c$ and $\xi_V$. The
reduced $\chi^2$ value is defined as $\chi^2\equiv (1/N)
\sum_{j=1}^N(I^{\rm experiment}_j-I^{\rm simulation}_j)^2/\sigma_j^2$
where $N$ is the number of pixels used in the fit, $I_j$ is the
intensity of pixel $j$ in the image and $\sigma_j$ is the noise
associated with pixel $j$. We used $\sigma_j^2=cI^{\rm simulation}_j$
having previously taken a series of images of the vacuum with
different electron intensities. A graph of the standard deviation
versus the average intensity showed the noise was Shot noise (so that
the square of the noise was proportional to the image intensity) and
gave the proportionality constant $c$ relating the counts recorded on
the detector to the number of electrons received. Unlike our previous
publication where separate fits were made for each
vortex~\cite{Loudon12}, here all the vortex images were fit
simultaneously with a single value of $\Lambda_{ab}$, $\Lambda_c$ and
$\xi_V$ using the model described in section~\ref{simulation}.


\section{Results}
\label{results}

\begin{figure}
\includegraphics[width=72mm]{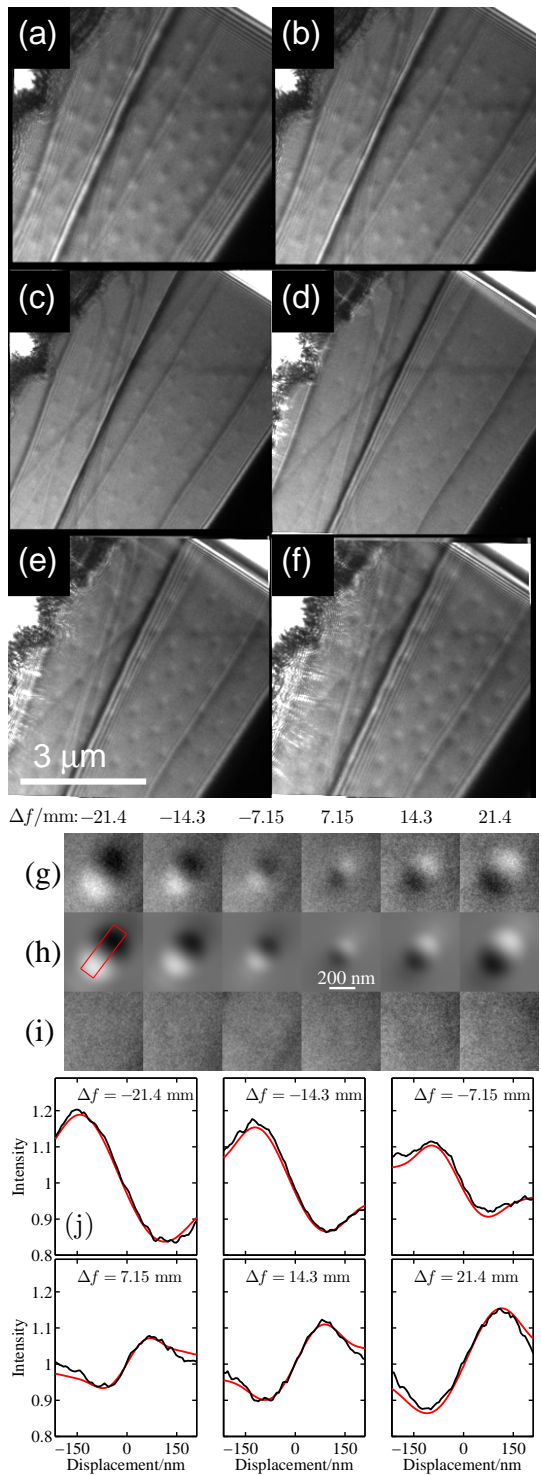}
\caption{\label{defseries} (a)--(f) Defocus series showing flux
  vortices in MgB$_2$ at 10.8~K in a field of 4.8~mT with defoci: (a)
  $\Delta f = -21.4$, (b) $-14.3$, (c) $-7.15$, (d) $7.15$, (e)
  $14.3$ and (f) $21.4$~mm. (g) Average vortex images at each defocus
  level. (h) Average simulated images. (i) Difference images between
  experiment and simulation. (j) Linescans across the images from the
  region shown by the red box in (h). Black lines show experimental
  data and red lines the fit.}
\end{figure}

Fig.~\ref{defseries}(a)--(f) show an experimental defocus series
acquired at 10.8~K (the base temperature of our cooling stage) in a
field of 4.8~mT. $\nv$ images of vortices which were least affected by
bend contrast were fit and with a single value of $\Lambda_{ab}$,
$\Lambda_c$ and $\xi_V$ along with the position of each vortex and its
in-plane rotation angle using a simplex algorithm~\cite{Press92} to
minimise the reduced $\chi^2$ value. The specimen thickness and tilt
angle were fixed at their calibrated values.

Fig.~\ref{defseries}(g) shows the average of these images at each
defocus level and (h) shows the average of the fitted simulations. To
demonstrate that the fit is good, (i) shows the difference between
images and simulations and (j) compares linescans taken across the
vortex images.

\begin{figure}
\includegraphics[width=70mm]{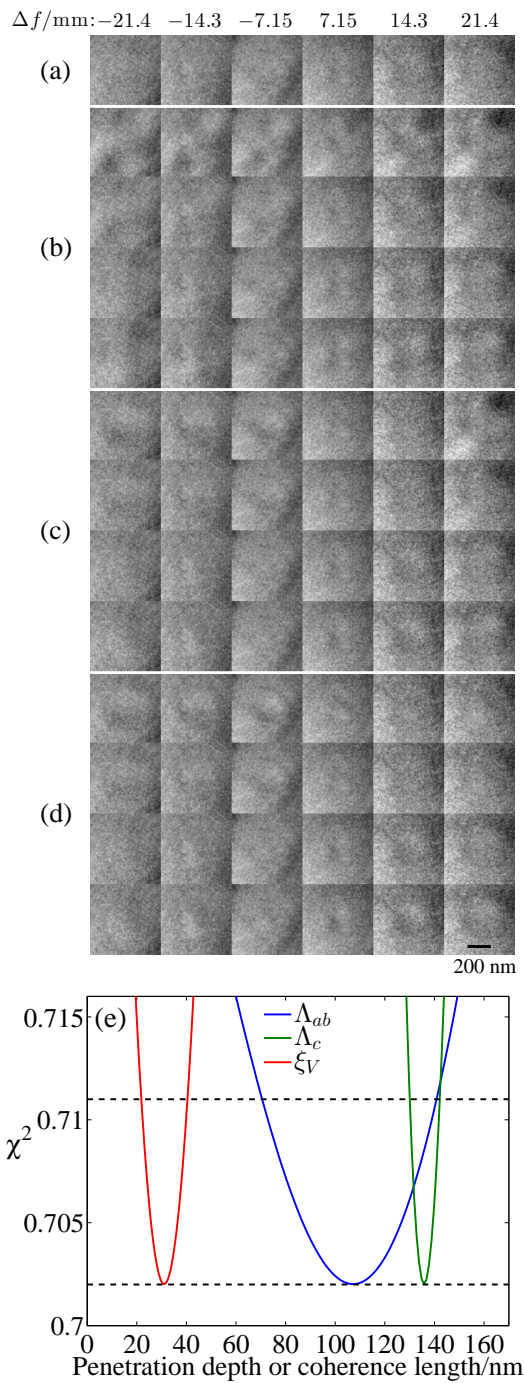}
\caption{\label{chi2} Difference images between the average defocus
  series and simulated images as each parameter is varied. (a)
  Difference images for the best-fit values of the parameters. (b)
  Difference images when $\Lambda_{ab}$ is varied: the top row shows
  difference images when $\Lambda_{ab}$ is set at two errorbars below
  the best-fit value. The next row is for $\Lambda_{ab}$ set one
  errorbar below. The next two rows are for $\Lambda_{ab}$ increased
  one and two errorbars above the best-fit value respectively. (c)
  A similar series showing the effect of changing $\Lambda_c$. (d)
  Series showing the effect of changing $\xi_V$. (e) Reduced $\chi^2$
  values as $\Lambda_{ab}$, $\Lambda_c$ and $\xi_V$ are
  adjusted. Acceptable values of $\chi^2$ lie between the dashed
  lines.}
\end{figure}

$\Lambda_{ab}$, $\Lambda_c$ and $\xi_V$ were then altered in turn and
the errorbar on each judged by the point at which the difference
images displayed a discernibly worse fit as shown in
Fig.~\ref{chi2}(a)--(d). This corresponded to an increase in the
reduced $\chi^2$ of 0.009 and the variation of $\chi^2$ as each
parameter is varies is shown in Fig.~\ref{chi2}(e). This yielded
$\Lambda_{ab}=\Lambdaafit\pm\Lambdaafite$~nm,
$\Lambda_c=\Lambdacfit\pm\Lambdacfite$~nm and
$\xi_V=\xiVfit\pm\xiVfite$~nm. The images are much more sensitive to
the value of $\Lambda_c$ than to $\Lambda_{ab}$ as a consequence of
mounting the sample tilted about the $a$ axis: had it been tilted
about $c$, the errors on $\Lambda_{ab}$ and $\Lambda_c$ would be
reversed. In a previous paper~\cite{Loudon13}, we obtained the more
precise value of $\Lambda_{ab}=107\pm 8$~nm using a sample cut in the
$ab$ plane.

To check for amorphous `dead-layers' of non-superconducting material
on the sample surfaces caused by ion thinning, the thickness of the
crystalline component of the sample was measured using the convergent
beam diffraction technique described in
ref.~\onlinecite{Williams96}. This showed no difference between the
total and crystalline thicknesses to within the experimental error of
$\pm 10$~nm. Reducing the calibrated thickness values by 50~nm to
account for the largest conceivable dead-layer reduced $\Lambda_{ab}$
by 13~nm, reduced $\Lambda_c$ by 27~nm and increased $\xi_V$ by 10~nm
without altering the quality of the fit. Taking this into account
gives: $\Lambda_{ab}=\Lambdaa\pm\Lambdaae$~nm,
$\Lambda_c=\Lambdac\pm\Lambdace$~nm and $\xi_V=\xiV\pm\xiVe$~nm. The
anisotropy ratio in the penetration depth is then
$\gamma_\Lambda=\gammaL\pm\gammaLe$ and the coherence lengths are
$\xi_{ab}=\gamma^{2/3}\xi_V=\xia\pm\xiae$~nm and
$\xi_c=\gamma^{-1/3}\xi_V=\xic\pm\xice$~nm. Alternatively, using the
more precise value of $\Lambda_{ab}=107\pm 8$~nm and
$\Lambda_c=\Lambdac\pm\Lambdace$~nm gives $\gamma_\Lambda=1.12\pm
0.16$, $\xi_{ab}=39\pm 11$~nm and $\xi_c=35\pm 10$~nm.

\section{Discussion}
\label{discussion}

We now compare the values obtained here with those found using other
techniques. These show that the conditions used (4.8~mT and 10.8~K)
were in the low field limit ($<100$~mT as established by neutron
diffraction) but not quite in the low temperature limit ($<5$~K from
radio-frequency measurements).

In 2005, Fletcher {\it et al.}~\cite{Fletcher05} performed
radio-frequency measurements which gave the change in the low-field
penetration depth with temperature but not absolute values. At 10.8~K,
$\Lambda_{ab}$ increased by $12\pm 1 $~nm and $\Lambda_c$ increased by
$18\pm 4 $~nm with respect to their low temperature
values. Subtracting these from our most precise measurements of the
penetration depths gives $\Lambda_{ab}=95 \pm\ 8$~nm and
$\Lambda_c=\Lambdaclow\pm\Lambdaclowe$~nm and the anisotropy as
$\gamma=1.07\pm 0.18$ in the low field and low temperature limit.

The most reliable measurement of the absolute value of the penetration
depth is likely to be from neutron diffraction and in 2003 Cubitt {\it
  et al.}~\cite{Cubitt03} found that at 2~K, the extrapolated low
field ($<100$~mT) value was $\Lambda_{ab}=82\pm 2$~nm which is close
to our value.

As samples grow as thin plates in the $ab$ plane, Cubitt {\it et al.}
did not have direct access to $\Lambda_c$ and so acquired diffraction
patterns with vortices tilted at $45^\circ$ with respect to the
$c$-axis. It was uncertain whether the formula used to calculate the
anisotropy was valid for a 2-band superconductor~\cite{Cubitt03} but
data acquired at 2~K between 0.2--0.5~T indicated that
$\gamma_\Lambda$ varied with field and its extrapolated value at low
field was $\gamma_\Lambda=1.1\pm 0.3$. In 2006, Pal {\it et
  al.}~\cite{Pal06} used a different neutron diffraction technique to
give $\gamma_\Lambda=1.1\pm 0.1$ at 4.9~K. Combining this with the
neutron value for $\Lambda_{ab}$ gives $\Lambda_c=90\pm 8$~nm which
agrees with our value of
$\Lambda_c=\Lambdaclow\pm\Lambdaclowe$~nm. The anisotropy we obtain is
close to the value of 1.01 calculated from first-principles by Golubov
{\it et al.}~\cite{Golubov02} in the clean limit but in common with
Fletcher {\it et al.}~\cite{Fletcher05}, we find penetration depths
approximately twice as large as predicted.

Cubitt {\it et al.} interpreted their data assuming that the coherence
length did not vary with field. If it did, the value they obtained,
$\xi_{ab}=8\pm 1$~nm, would apply only at high field ($>0.8$~T). This is
close to the value of 10~nm found from the upper critical
field~\cite{Eskildsen03b}.

Eskildsen {\it et al.}~\cite{Eskildsen02} measured the coherence
length in 2002 using scanning tunnelling microscopy (STM) to measure
the width of vortex cores, scanning the $ab$ plane with tunnelling in
$c$. As the $\sigma$ carriers are confined to the $ab$-planes, the
tunnelling current came almost exclusively from the $\pi$ band giving
$\xi_\pi=38.8\pm 0.7$~nm at 0.32~K in a field of 50~mT (after
adjusting for their slightly different model for the core). This agrees
with our value of $\xi_c=35\pm 10$~nm at 10.8~K and 4.8~mT and
supports this larger value at low field.

\section{Summary and Conclusions}
\label{conclusions}

We have described a new method to measure the penetration depth and
coherence length of a superconductor in all directions at low applied
magnetic field using transmission electron microscopy. The measurement
does not need large-scale facilities and required one day to thin and
mount the sample and another day to take the images required. The
experiment was performed on a very small sample: $30\times
15$~$\upmu$m and 200~nm thick so this method could prove useful for
superconductors where only very small single crystals are available,
as is the case for some iron-based superconductors. It is also useful
if the penetration depth and coherence length vary with field, as is
the case for MgB$_2$, as the measurement is made at very low fields
which can be difficult to access using other techniques.

For a sample of MgB$_2$ cut in the $ac$ plane and tilted to $45\pm
5^\circ$ about the $a$-axis, we obtained
$\Lambda_{ab}=\Lambdaa\pm\Lambdaae$~nm,
$\Lambda_c=\Lambdac\pm\Lambdace$~nm at 10.8~K in a field of
4.8~mT. The large error in $\Lambda_{ab}$ is a consequence of tilting
the sample about the crystallographic $a$ axis. Had it been tilted
about $c$ instead, the errors on $\Lambda_{ab}$ and $\Lambda_c$ would
be reversed. We obtained a more precise value of $\Lambda_{ab}=107\pm
8$~nm at 10.8~K in our previous paper~\cite{Loudon13} in which the
sample was cut in the $ab$ plane. Using this value gives
$\Lambda_{ab}=107\pm 8$~nm, $\Lambda_c=\Lambdac\pm\Lambdace$~nm,
$\xi_{ab}=39\pm 11$~nm and $\xi_c=35\pm 10$~nm which agree well with
measurements made using other techniques discussed in
section~\ref{discussion}.

Obtaining the most precise values for the penetration depths and
coherence lengths in all directions using this technique requires
taking images of the vortex lattice with the sample tilted in more
than one direction. It might be thought that the sample could be
mounted in the plane of the support grid and tilted with the
microscope goniometer, first about $a$ and then about $c$. However,
the design of conventional liquid helium cooled holders does not allow
tilting to an angle higher than $\alpha=25^\circ$, so the sample was
instead mounted to the support grid at $45\pm 5^\circ$ to give
sufficient contrast in the image. Thus, to investigate a new
superconductor and obtain the most accurate measurement of the
penetration depth and coherence length in all directions, it would be
best to cut two samples and mount both to the grid, one tilted about
$a$ and the other tilted about $c$.

We used the Ginzburg-Landau model for the magnetic structure of flux
vortices but as MgB$_2$ is a two-band superconductor, the vortices may
well have a more complex structure as described in
ref.~\onlinecite{Babaev10}. We obtained good fits to the vortex images
and there was no indication of a more complex structure within the
accuracy of the technique but our simulation scheme would allow any
model for the vortex structure to be used for image simulations
provided that the magnetic vector potential for a vortex in an
infinitely thick superconductor is known.


\begin{acknowledgments}
This work was funded by the Royal Society (United Kingdom). Work at
Eidgen{\"o}ssische Technische Hochschule, Z{\"u}rich was supported by
the Swiss National Science Foundation (Switzerland) and the National
Center of Competence in Research programme `Materials with Novel
Electronic Properties'.
\end{acknowledgments}


\bibliography{MgB2}

\end{document}